\documentclass[aps,prl,superscriptaddress,twocolumn,amsmath,amssymb,floatfix]{revtex4-1}

\usepackage{pdfpages}
\makeatletter
\AtBeginDocument{\let\LS@rot\@undefined}
\makeatother

\usepackage{dcolumn}% Align table columns on decimal point
\usepackage{bm}% bold math
\usepackage{microtype}
\usepackage{braket}

\usepackage{graphicx}% Include figure files
\usepackage{hyperref}
\hypersetup{colorlinks = true, citecolor = blue, linkcolor = red}

\usepackage[caption=false]{subfig}
\usepackage{float}

\usepackage{braket}
\usepackage{dsfont}
\usepackage{commath}

\usepackage[utf8]{inputenc}
\usepackage[T1]{fontenc}

\renewcommand{\i}{\mathrm{i}} % Upright i for iota

\begin{document}

\title{Proposal for Quantum Simulation via All-Optically Generated Tensor Network States}% Force line %\thanks{A footnote to the article title}%
\author{I.~Dhand}
%\email{ishdhand@gmail.com}
\affiliation{Institut f{\"u}r Theoretische Physik, Albert-Einstein-Allee 11, Universit{\"a}t Ulm, 89069 Ulm, Germany}
\affiliation{Center for Integrated Quantum Science and Technology (IQST), Albert-Einstein-Allee 11, Universit{\"a}t Ulm, 89069 Ulm, Germany}

\author{M.~Engelkemeier}
\affiliation{Department of Physics and CeOPP, University of Paderborn, Warburger Strasse 100, D-33098 Paderborn, Germany}

\author{L.~Sansoni}
\affiliation{Department of Physics and CeOPP, University of Paderborn, Warburger Strasse 100, D-33098 Paderborn, Germany}

\author{S.~Barkhofen}
\affiliation{Department of Physics and CeOPP, University of Paderborn, Warburger Strasse 100, D-33098 Paderborn, Germany}

\author{C.~Silberhorn}
\affiliation{Department of Physics and CeOPP, University of Paderborn, Warburger Strasse 100, D-33098 Paderborn, Germany}

\author{M.~B.~Plenio}
\affiliation{Institut f{\"u}r Theoretische Physik, Albert-Einstein-Allee 11, Universit{\"a}t Ulm, 89069 Ulm, Germany}
\affiliation{Center for Integrated Quantum Science and Technology (IQST), Albert-Einstein-Allee 11, Universit{\"a}t Ulm, 89069 Ulm, Germany}

\date{\today}
\begin{abstract}
We devise an all-optical scheme for the generation of entangled multimode photonic states encoded in temporal modes of light.
The scheme employs a nonlinear down-conversion process in an optical loop to generate one- and higher-dimensional tensor network states of light.
We illustrate the principle with the generation of two different classes of entangled tensor network states and report on a variational algorithm to simulate the ground-state physics of many-body systems.
We demonstrate that state-of-the-art optical devices are capable of determining the ground-state properties of the spin-$1/2$ Heisenberg model.
Finally, implementations of the scheme are demonstrated to be robust against realistic losses and mode mismatch.
\end{abstract}

\maketitle

General quantum states possess a complex entanglement structure that makes their description on a classical computer inefficient in the sense that, generally, the computational effort grows exponentially with the number of subsystems. 
However, in ground and thermal states of local Hamiltonians the entanglement and correlations are typically more limited as they satisfy area laws~\cite{Audenaert2002,Eisert2010,Brandao2015}.
Such states can be approximated well in terms of matrix product states (MPSs) or, more generally, tensor network (TN) states parametrization, in which only a polynomial, in the number of subsystems, number of parameters is required to describe the state~\cite{Orus2014,Verstraete2008}. 
This class includes not only the ground states of a wide variety of quantum many-body Hamiltonians~\cite{Eisert2010,Schuch2008a} but also eponymous examples of entangled states such as the Greenberger-Horne-Zeilinger (GHZ) state and $W$ state. 
The generation of MPSs is important as they include important resource states for quantum communication, teleportation and metrology~\cite{Wang2005,Cao2005,Wang2007,Jarzyna2013}.
Furthermore, TNs can efficiently parametrize important quantum states including universal states for quantum computation (e.g., cluster and Affleck-Kennedy-Lieb-Tasaki states~\cite{Raussendorf2001,Wei2011}), states important in high-$T_{\text{c}}$ superconductivity (e.g., resonating valence bond state~\cite{Anderson1987}), and topologically ordered states of matter~\cite{Kitaev2003,Verstraete2006}.
Although matrix product states can be efficiently manipulated on a classical computer~\cite{Schollwoeck2011}, the treatment of TN states in higher spatial dimensions remains challenging because the computational effort, while polynomial, grows with a high power in the number of subsystems and bond dimension.
Therefore, the experimental generation of TN states and their use for quantum simulation is of considerable interest.

\begin{figure}
\includegraphics[width=\columnwidth]{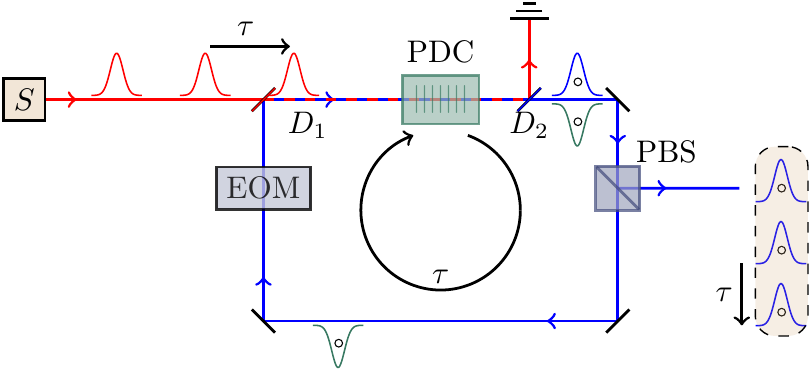}
\caption{Setup to generate 1D TN states: 
the setup includes a PDC nonlinearity placed inside an optical loop. 
Laser pulses (red) from a source $S$ placed outside of the loop are fed into it, where they pump the nonlinearity to effect PDC on the two polarization modes of light.
One polarization mode (blue) is coupled out of the loop using a PBS. 
The cycling time $\tau$ of the signal (green) mode equals the time separation of the pump pulses. 
Dichroic mirrors ${D}_{1}$ and  ${D}_{2}$ couple the pump out.
The circles represent superpositions over low-photon-number Fock states.
The light modes coupled out from the loop via the PBS contain the desired 1D TN state.
}
\label{Fig:Setup1D}
\end{figure}

Current experimental implementations for the generation and processing of TN states focus on spatial modes of light, but these implementations require experimental resources that typically increase quickly with the required size of the TN state~\cite{Zou2002,Zou2005,Su2007,Yukawa2008,Huang2011}.
This limitation can be overcome by using the temporal modes of light or time bins, which provide an infinite-dimensional Hilbert space that can be controlled with constant experimental resources through time multiplexing. 
The potential of this approach has already been successfully demonstrated in the context of quantum walks and boson sampling~\cite{Schreiber2010,Schreiber2011,Motes2014,He2017}. 
Existing proposals for generating photonic TN states in the temporal modes of light rely on the strong coupling of light to a single atom trapped inside a cavity~\cite{Schon2005,Schon2007,Pichler2017}.
The strength of these methods is that they allow the generation of arbitrary 1D TN states whose entanglement is limited only by the number of accessible atomic states.
However, the experimental implementation of these schemes requires two challenging conditions to be met, namely, the cooling and localizing of the atom, and strong coupling between the atom and the light emitted from the cavity.
Moreover, the requirement of complete control over multiple atomic states restricts the amount of entanglement in the generated TN states.

In this Letter, we devise an all-optical scheme for the generation of TN states in one and higher dimensions that overcomes these challenges.
Our scheme does not suffer from the stringent requirement of strong atom-photon coupling and instead exploits well established parametric down-conversion (PDC) methods to build entanglement in the generated state~\cite{Wu1986}.
Furthermore, our method overcomes the restriction on entanglement (as quantified by bond dimension) to accessible atomic levels by using the photon-number degree of freedom to share entanglement between components of the generated state. 
Finally, our all-optical scheme also promises robustness against loss and mode mismatch and can be realized with current optical technology.

\textit{Scheme to generate TN states.} --- 
Our proposed scheme to generate entangled multimode states of light is depicted in Fig.~\ref{Fig:Setup1D}.
The experimental setup relies on placing a type-II PDC nonlinearity into an optical loop and optically pumping the nonlinearity.
This nonlinearity performs two-mode squeezing
$U = \exp(\eta \hat{a}_\text{h}^{\dagger}\hat{a}_\text{v}^{\dagger} - \eta^{*}\hat{a}_\text{h}\hat{a}_\text{v})$ on the horizontal and vertical modes of light, where the PDC parameter $\eta$ depends on the strength of the optical pumping.
Here $\hat{a}_{i}^\dagger$ and $\hat{a}_{i}$ are the creation and annihilation operators for mode $i\in\{\text{h}, \text{v}\}$.
The light in one of the two polarization modes (say, vertical) is coupled out of the loop via a polarizing beamsplitter (PBS), while the other (say, horizontal) cycles the loop.
An electro-optic modulator (EOM) in the loop dynamically mixes the two polarization modes, of which the vertical mode is in vacuum, via arbitrary linear transformations 
$\hat{a}^{\dagger}_{j} \to \sum_{i = 1}^{2}V_{ij} \hat{a}^{\dagger}_{i}$ for $2\times 2$ special unitary matrix~$V\in \text{SU}(2)$~\cite{Capmany2011}.
Sec.~A of the Supplementary Material~\cite{sm} details the modeling of the setup.
The time it takes for light to cycle the loop is set equal to the delay between subsequent pump pulses.
Thus, the cycling light arrives synchronous to the next pump pulse and effects two-mode squeezing interaction between the two polarization modes~\cite{Simon2000}.
 In other words, the PDC and the EOM together give rise to an interaction between the horizontally polarized cycling light and the vertically polarized optical vacuum.

\begin{figure}
\includegraphics[width=\columnwidth]{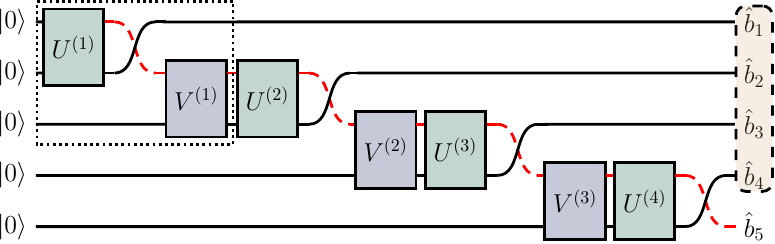}
\caption{Quantum circuit of 1D experiment for four cycles.
$U^{(i)}$ represents the two-mode PDC process and $V^{(i)}$ represents the EOM transformation in the $i$th cycle.
The dotted box represents one cycle.
The action of the PBS is represented by the mode swapping after each $U^{(i)}$ operation and a subsequent emission of one of the polarization modes.
The red dashed line represents the cycling mode.
The rounded box on the right encloses subsequently emitted temporal modes $\{\hat{b}_{j}\}$ that contain the state of interest.}
\label{Fig:Circuit}
\end{figure}

The quantum circuit representing this repeated interaction is presented in Fig.~\ref{Fig:Circuit}.
We consider the temporal modes (represented by $\{\hat{b}_{j}, \hat{b}_{j}^{\dagger}\}$) of the light coupled out from the loop over many cycles, where each temporal mode is the vertically polarized mode that was coupled out from the PBS at a different time.
We show the establishment of multi-particle entanglement between subsequent temporal modes mediated by the light cycling in the loop as depicted by the dashed line of Fig~\ref{Fig:Circuit}.
Specifically, we show that the emitted temporal modes of light permit a 1D TN representation and include entangled states such as $W$ and GHZ states.
The proof for this result and the general form of the resultant TN state is in Supplementary Material~Sec.~B~\cite{sm}. 
The intuition for the proof is that the cycling mode mediates entanglement between subsequently emitted light modes.
Entanglement between one emitted mode and the next is limited by the entanglement between the first mode and the cycling mode, and this maximum entanglement is constant irrespective of the number of cycles. 
Because subsequent temporal modes of light are entangled, albeit with limited entanglement, it follows that the state of the emitted light can be represented as a TN state of limited bond dimension.

\begin{figure}
\centering
\includegraphics[width=\columnwidth]{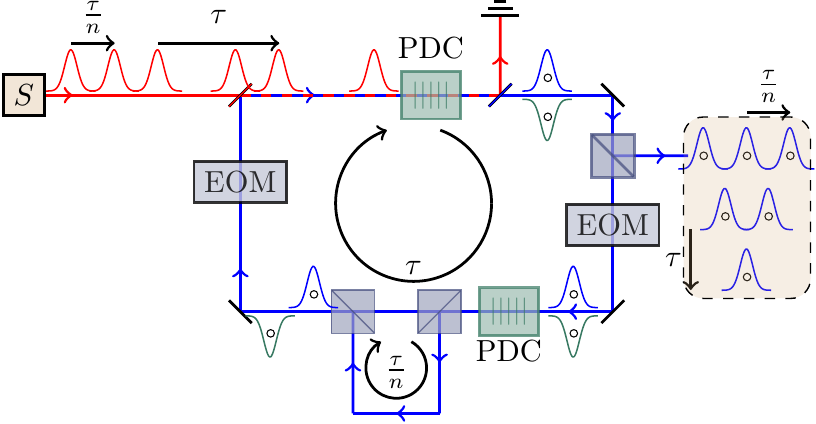}
\caption{\textbf{Setup to generate 2D TN states}: 
to generate 2D TN states, an additional fibre loop (corresponding to a time delay $\tau/n$ for chosen integer $n$) is connected into the existing 1D loop via PBSs.
Pumping of the additional optional PDC is omitted from the figure for simplicity.
TN states in more than two dimensions can be generated by introducing additional fiber loops into the optical setup.
}
\label{Fig:Setup2D}
\end{figure}

Although the properties of 1D TN states can be efficiently obtained on a classical computer, those of TN states in two and higher dimensions require classical algorithms that scale badly, i.e., exponentially in the system size and as high-degree polynomials in the bond dimension. 
In other words, two- and higher-dimensional TN states can be exploited for obtaining nontrivial quantum-computational speed up.
It is possible to modify our scheme to generate higher-dimensional TN states, by connecting additional optical loops into the existing loop as depicted in Fig.~\ref{Fig:Setup2D}.
The effect of one additional loop is to convert different polarization modes into temporal modes, an approach already used in 2D quantum walks~\cite{Schreiber2010,Schreiber2012}.
Optionally, additional nonlinearities and EOMs can be added to the loop to ensure that the entanglement structure is identical in the two dimensions of the lattice.
The additional optical loop is designed to provide a time delay of $\tau/n$, which is smaller than the cycling time $\tau$ of the main loop by a factor $n$ for some large integer $n$.
Owing to this additional time delay, the difference between the emission times of two temporal modes is either $\tau$ or multiples $\tau/n$, $2\tau/n$, $3\tau/n,\dots$ of the interval $\tau/n$.
Modes with time difference $\tau$ are interpreted as neighbors along one axis of the TN lattice, whereas those with time difference $\tau/n$ are interpreted as neighbors along a different axis.
Depending on the required number $\tilde{n}$ of lattice site along the second TN dimension, we can choose any $n > \tilde{n}$ so that the sites in the 2D lattice are uniquely defined.
Thus, the emitted light possesses an entanglement structure that is captured by a 2D TN state with a triangular structure (See Supplementary Material Sec.~B~\cite{sm}).
Similarly, additional loops can be connected to the optical setup to generate higher-dimensional TN states.
We estimate that current optical technology can enable the generation of five-mode 1D TN states and 15-mode 2D TN states at the rates of 1200 and 85\,Hz, respectively (Supplementary Material Sec.~E~\cite{sm}).

\textit{State generation.} ---
Here we detail how the setup can be used to generate two inequivalent classes of entangled states, namely the $W$ state and the GHZ state.
First, we consider the $m$-qubit $W$ state $\ket{\psi}_{W} = \ket{0\dots 01} + \ket{0\dots 10}+ \dots + \ket{1\dots00}$,
which has one excitation $\ket{1}$ that is delocalized uniformly over all the qubits. 
Our proposed setup can generate a heralded $W$ state, which is defined as
\begin{align}
\ket{\psi}_{HW} =\,& \ket{0\dots 00}\otimes\ket{0} +\eta\big( \ket{0\dots 01}\nonumber \\
& +\ket{0\dots 10}+ \dots + \ket{1\dots00}\big) \otimes \ket{1}
\label{Eq:HeraldedW}
\end{align}
on a total of $m +1$ qubits for some complex $\eta$ with $\eta < 1$ and the normalization factor is emitted for simplicity.
In this state, a $\ket{1}$ in the last qubit heralds the presence of a W state in the remaining qubits whereas a $\ket{0}$ in the last qubit implies a vacuum state in the remaining qubits.

The heralded W state can be generated by our proposed setup in the single-rail basis~\cite{Ralph2010}, wherein the absence of a photon in a temporal mode encodes the state $\ket{0}$ and a single photon in the mode encodes $\ket{1}$.
Cases where more than a single photon is present in the mode are discarded.
Even after accounting for this postselection, high rates of state generation of the order of kilohertz can be obtained (Supplementary Material Sec.~E~\cite{sm}).

Next we describe the generation of the 4-qubit GHZ state~\cite{Greenberger1989}, which is usually defined as an equal superposition $\ket{0000} + \ket{1111}$ over each qubit that is in state $\ket{0}$ and each qubit in state $\ket{1}$.
An alternative description of the GHZ state $\ket{1100} + \ket{0011}$ is obtained by redefining the qubit labels in the last two qubits.
Our proposed setup can be used to generate the diluted GHZ state 
\begin{equation}
\ket{\psi}_{\text{GHZ}}=\, 
\ket{0000} + \eta \left(\ket{0011} + \ket{1100}\right),
\label{Eq:FourModeGHZ}
\end{equation}
where the normalization factor is omitted.
Supplementary Material Sec.~C details the optical circuit parameters for generating these states~\cite{sm}.

\begin{figure}
\includegraphics[width = 1\columnwidth]{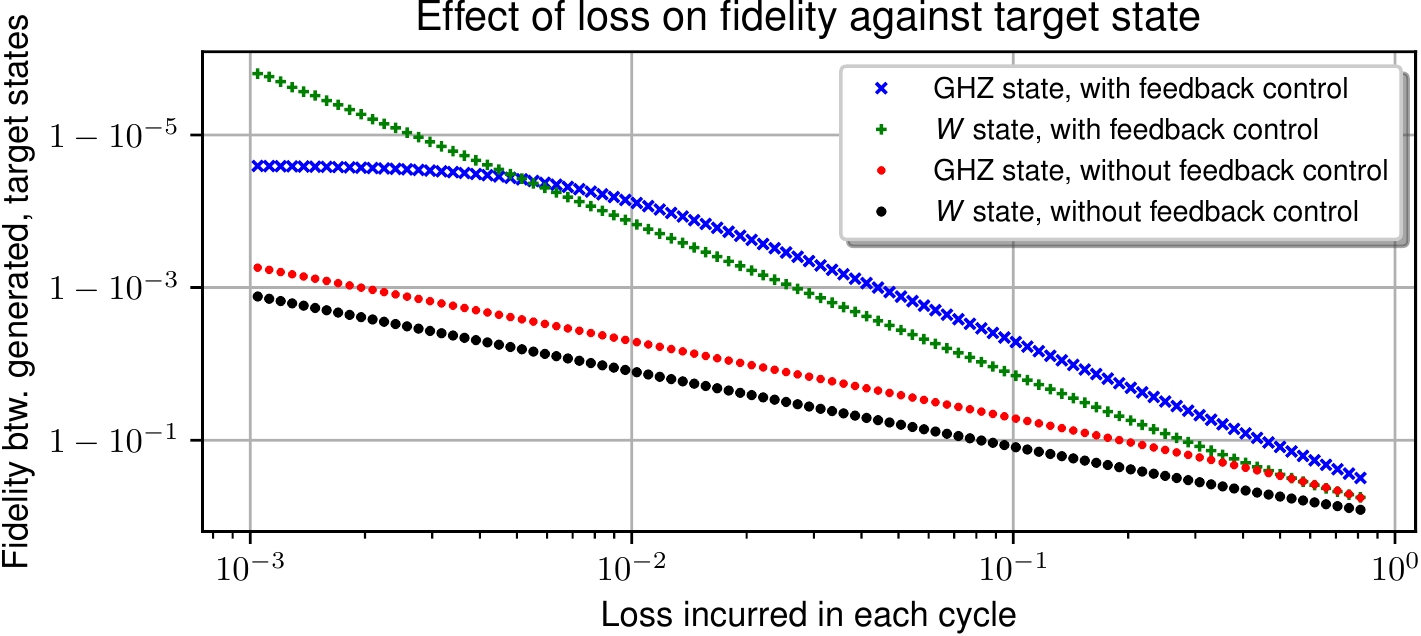}\label{Fig:LossW}
\caption{Simulations: effect of loss on the fidelity of the generated four-qubit state with respect to target W and GHZ states.
The red and black dots represent the fidelity between the respective generated and target state as a function of the loss incurred by the light in each cycle.
The blue and green crosses represent the same quantity under self-correction, i.e., when variational algorithms are used to find circuit parameters that optimize the fidelity against the target state in the lossy case.}
%See Supplementary Material Sections~D and~E for details on simulation and experimental considerations respectively.
\label{Fig:W_and_GHZ}
\end{figure}

In both cases of heralded $W$- and diluted GHZ-state generation, we can obtain experimental results for $W$- and GHZ-states by post-selecting only those experimental outcomes in which the expected numbers of photons were observed.
Simulations provide evidence that our state generation procedure is robust against the usual experimental imperfections of loss and mode mismatch from the PDC.
Consider reasonable experimental losses which are typically upward of 10\% loss in each cycle; these losses can lead to higher than 90\% fidelity with respect to target state as seen in the red and black dots of Fig.~\ref{Fig:W_and_GHZ}.

\textit{Quantum-variational algorithm.} --- 
Other than state preparation, the proposed setup can be exploited for performing a mixed quantum-classical algorithm for the determination of ground state properties of many-body systems via a quantum-variational approach, which we now describe.
We consider the task of determining the properties, such as the energy or correlations, of the ground state of a given Hamiltonian operator that acts on qubits.
The generated TN states comprise the set of variational states; their energy with respect to the given Hamiltonian is obtained by performing Glauber correlation measurements on the output light following the procedure of~\cite{Barrett2013} in single-rail representation~\cite{Lund2002,Donati2014}.
A classical minimization algorithm can then be used to obtain circuit parameters corresponding to the generated state that has the lowest energy with respect to the given Hamiltonian. 
If the circuit parameters, including the pump strength and EOM parameters, are sufficiently expressive, i.e., if the ground state is close to the class of variational states generated by the setup, then an accurate approximation of the given Hamiltonian's ground state can be obtained. 
The procedure is expected to work well for a wide variety of Hamiltonians because the ground state of most 1D local Hamiltonians is close to a low-dimension TN state~\cite{Eisert2010}.
The properties of the ground state can be determined by usual measurements on the output light.
Our mixed quantum-classical variational approach encompasses the variational problem that can be solved using the so-called Ising machines because it exploits the polarization and photon-number degrees of freedom in addition to temporal modes used in Ising machines~\cite{McMahon2016,Clements2017}.

\begin{figure}
\subfloat[]{\includegraphics[width=\columnwidth]{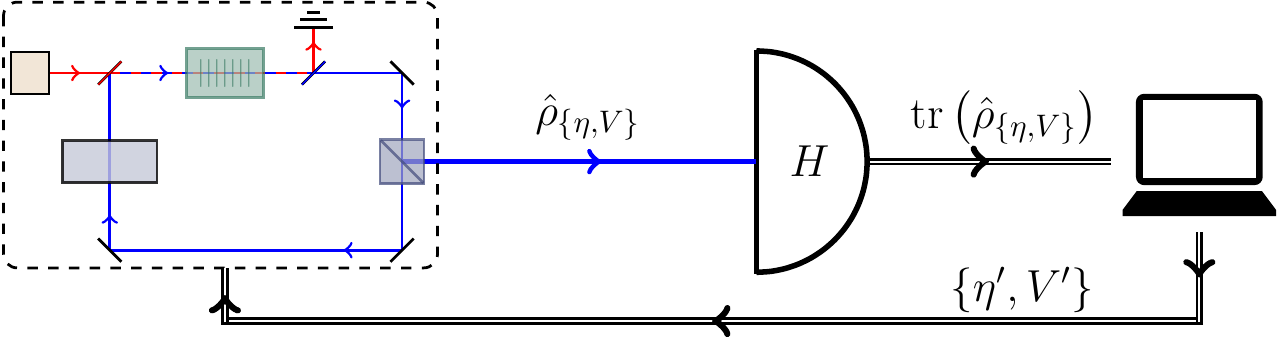}\label{Fig:Flowchart}}\\
\subfloat[]{\includegraphics[width = 0.97 \columnwidth]{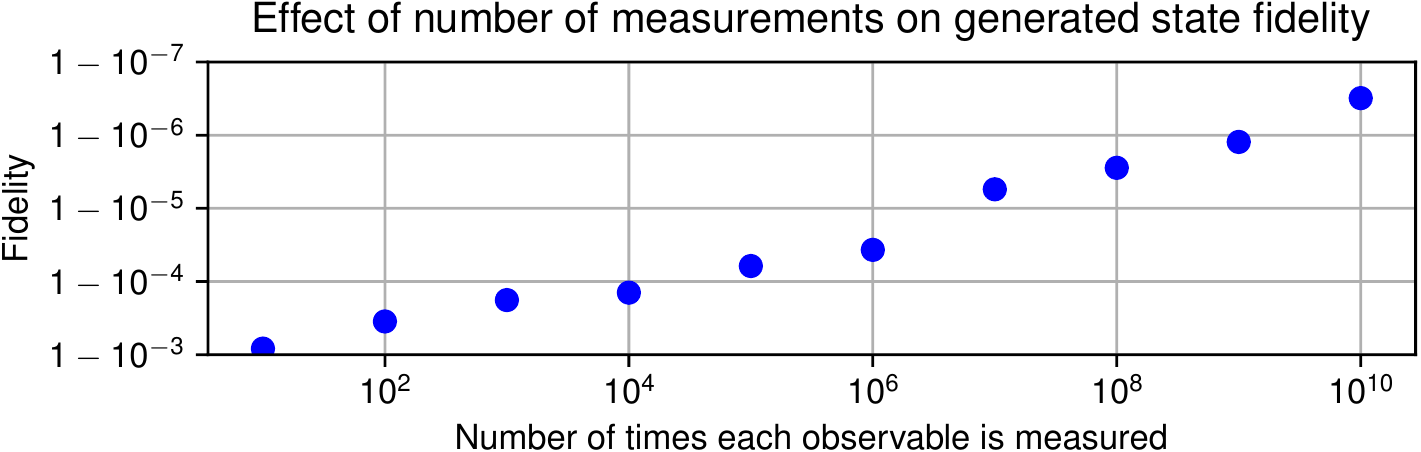}\label{Fig:QVAFinite}}\\
\subfloat[]{\includegraphics[width = 0.95 \columnwidth]{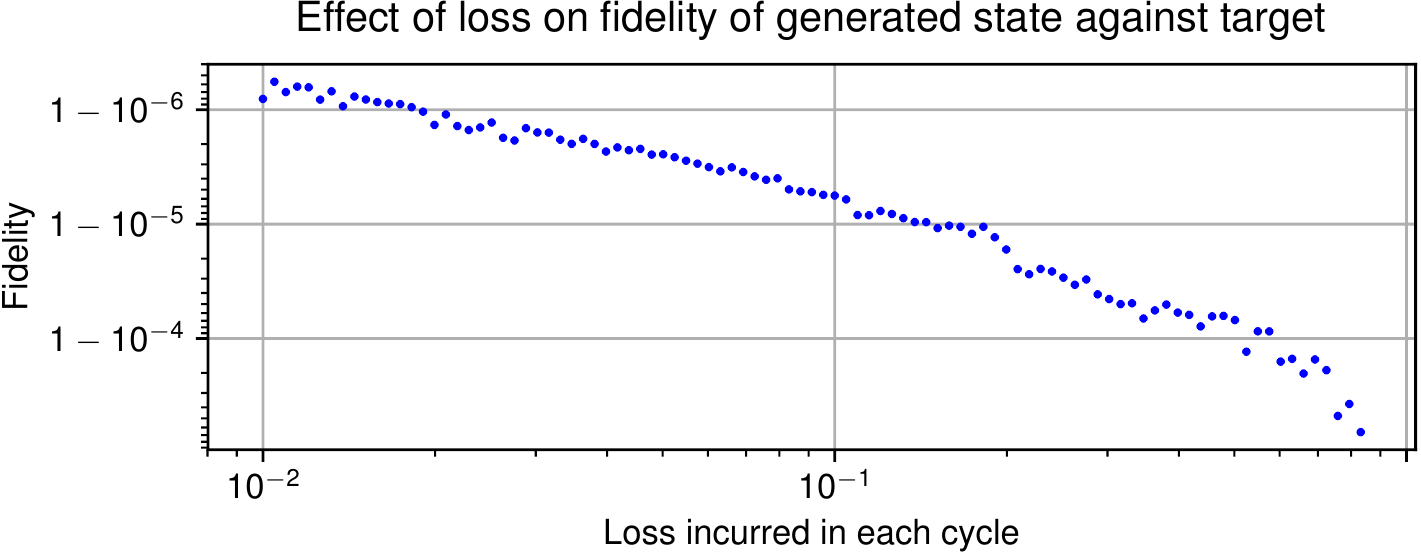}\label{Fig:QVALoss}}
\caption{\textbf{(a) Scheme of quantum-variational algorithm and simulated performance under (b) finite numbers of measurements and (c) losses.} 
(a) Depiction of quantum-variational algorithm.
The output from the setup (parameters set to $\{\eta\}, \{V\}$) is fed into detection setup that encodes given Hamiltonian  $H$.
The detector output is analyzed by a classical optimization routine to choose the set of variational parameters $\{\eta'\}, \{V'\}$ for the next step.
(b) The simulated number of measurements performed for each observable versus the fidelity $F$ between the expected ground state (W state) and the state obtained from the quantum-variational algorithm, without including the effect of losses.
(c) The fidelity between the expected ground state and the state obtained from the quantum-variational algorithms as a function of simulated loss in each loop.
The variational algorithm chooses a different pumping value for each cycle.
See Supplementary Material Sec.~D for simulation details and Sec.~E for experimental considerations~\cite{sm}.}
\label{Fig:QVA}
\end{figure}

To illustrate the performance of this approach, we simulate the procedure to find the ground state of the isotropic XY model~\cite{Lieb1961,Takahashi2005}.
The ground state of XY Hamiltonian
$ H_{\text{XY}} = J \sum_{i} X_{i}X_{i+1} + Y_{i}Y_{i+1} + \frac{B}{4}\sum_{i} Z_{i}$
is the $W$ state for certain range of $B$~\cite{Zhang2012}.
We simulate Glauber correlation measurements on the output light to obtain the energy of the generated state for a specific value of circuit parameters~\cite{Barrett2013}. 
Starting with random circuit parameters, we use a constrained minimization algorithm to find those circuit parameters that minimize the energy.
The variational minimization returns a state that is close to the expected ground state as depicted in Fig.~\ref{Fig:QVA}.
Simulations provide evidence that this approach is robust against statistical noise (Fig.~\ref{Fig:QVAFinite}) and loss (Fig.~\ref{Fig:QVALoss}).

A similar variational approach can also be used to enhance the quality of state generation, as described above, against possible experimental imperfections.
For instance, consider the task of improving the fidelity $F = \bra{\psi_{t}}\rho_{\text{lab}}\ket{\psi_{t}}$ of the generated state $\rho_{\text{lab}}$ with respect to a given target state $\ket{\psi_{t}}$, such as the W state, under the presence of imperfections such as loss and phase drift.
We can leverage from a measurement-based feedback control scheme~\cite{Inoue2013} to find the circuit parameters that maximize fidelity against a desired state and thereby compensate for experimental imperfections.
Direct fidelity estimation procedures~\cite{Flammia2011,Silva2011} can be used to efficiently estimate the fidelity with respect to the desired state and classical optimization can be performed to maximize this fidelity. 
Simulations show that our $W$ and GHZ state generation procedures can be made further resilient to loss via such feedback control by 2-3 orders of magnitude (see blue and green crosses in Fig.~\ref{Fig:W_and_GHZ}).

\textit{Discussion.} ---
In summary, we propose a scheme for the all-optical generation of one- and higher-dimensional TN states in temporal modes of light.
The free parameters describing the TN state and its bond dimension can be improved by using additional degrees of freedom of light, such as spatial modes, time-frequency Schmidt modes and orbital angular momentum modes of light~\cite{Reck1994,Marrucci2011,Brecht2015,Dhand2015},
or by adding another EOM to the loop between the PDC and the PBS.
Finally, states such as coherent states can be impinged into the PDC instead of starting with the optical vacuum, thereby leading to the generation of high-photon-number Gaussian matrix product states~\cite{Gerry1995,Schuch2008}, which could potentially be used as a resource for Gaussian boson sampling~\cite{Hamilton2017,Chakhmakhchyan2017}. 
Efficient TN-based procedures can be employed to perform tomography of the states~\cite{Cramer2010,Baumgratz2013,Baumgratz2013a,Lanyon2017}. 

\begin{acknowledgments}
We thank Raul Garcia-Patron, Sandeep K.~Goyal, Milan Holz\"apfel, W.~Steven Kolthammer, Leonardo Novo, and Johannes Tiedau for helpful discussions.
We acknowledge support by the state of Baden-W\"urttemberg through bwHPC and the German Research Foundation (DFG) through grant number INST 40/467-1 FUGG.
The group at Ulm was supported by the ERC Synergy grant BioQ, the EU project QUCHIP, and by the Alexander von Humboldt Foundation via the Humboldt Research Fellowship for Postdoctoral Researchers. 
The group at Paderborn acknowledges financial support from the Gottfried Wilhelm Leibniz-Preis (Grant No.~SI1115/3-1), from the European Union's Horizon 2020 research and innovation program under the QUCHIP project (Grant No.~641039) and from European Commission with the ERC project QuPoPCoRN (No.~725366).
\end{acknowledgments}

\emph{Note added.}---Please also see related article~\cite{Lubasch2017} by Lubasch~\emph{et al.}.


%\bibliography{QSim_all-optical_TN}
\clearpage


\section*{Supplementary material}

# Supplemental Material: Quantum Simulation via All-Optically-Generated Tensor Network States

I. Dhand,[1,2] M. Engelkemeier,[3] L. Sansoni,[3] S. Barkhofen,[3] C. Silberhorn,[3] and M. B. Plenio[1,2]

[1]*Institut für Theoretische Physik, Albert-Einstein-Allee 11, Universität Ulm, 89069 Ulm, Germany*
[2]*Center for Integrated Quantum Science and Technology (IQST),*
*Albert-Einstein-Allee 11, Universität Ulm, 89069 Ulm, Germany*
[3]*Department of Physics and CeOPP, University of Paderborn,*
*Warburger Strasse 100, D-33098 Paderborn, Germany*

## A. MODELING THE SETUP

Here we provide details about the modeling of the setup and the generation of $W$ and GHZ states as described in the main text. First, we describe our modeling of the PDC and the EOM in the setup. The PDC performs two-mode squeezing on the light in the two orthogonal polarization modes. We model the downconversion process by the Hamiltonian

$$\mathcal{H} = \int d\omega_1 \, d\omega_2 \, F(\omega_1, \omega_2) \left( \kappa \hat{a}_h^\dagger \hat{a}_v^\dagger + \kappa^* \hat{a}_h \hat{a}_v \right), \quad (1)$$

where $\hat{a}_h$ and $\hat{a}_v$ are the photon annihilation operators corresponding to the two polarization modes; the spectral amplitude $F(\omega_1, \omega_2)$ depends upon properties of the pump and the nonlinearity; and $\kappa$ is the complex-valued interaction strength [1, 2]. We assume no spectral correlations and perfect mode matching. While the latter of these two imperfections can be accounted in terms of additional loss in the loop, the former can be modeled by considering dilution with a fully-mixed state [3, 4]. Under these assumptions, the analysis simplifies to monochromatic photons and the Hamiltonian to

$$H = \kappa \hat{a}_h^\dagger \hat{a}_v^\dagger + \kappa^* \hat{a}_h \hat{a}_v. \quad (2)$$

Note that the interaction strength $\kappa$ depends on the pump power, which can be adjusted independently for each cycle. The unitary transformation effected by this Hamiltonian is given by

$$U = \exp(-iHt) \quad (3)$$
$$= \exp(\eta \hat{a}_h^\dagger \hat{a}_v^\dagger - \eta^* \hat{a}_h \hat{a}_v) \quad (4)$$

where the magnitude $|\eta|$ of the complex PDC parameter

$$\eta \stackrel{\text{def}}{=} -i\kappa t \quad (5)$$

expresses the strength of interaction between the two polarization modes.

The EOM placed in the optical loop allows changing the phase and the polarization of the incoming field with each cycle. Thus, the EOM effects a beamsplitter-like transformation

$$\hat{a}_j^\dagger \to \sum_{i=1}^{2} V_{ij} \hat{a}_i^\dagger \quad (6)$$

on the light in the two polarization modes for $2 \times 2$ unitary transformation $V$ [5]. The PDC parameter $\eta$ and the elements of the SU(2) EOM transformation matrix $V_{ij}$ comprise the set of free parameters in the generation of the state. The scheme is independent of the exact type or nonlinear material used in the implementation. Moreover, nonlinearities other than PDC can also be used as long as these involve an interaction between two orthogonally polarized modes of light.

The PBS couples one of the two polarization modes out of the loop. Specifically, the temporal mode $\hat{b}_j$ emitted from the PBS in the $j$-th cycle is the (say) vertical polarization mode $\hat{a}_v$. The horizontal polarization mode $\hat{a}_h$ stays cycling in the loop and acts as the first input to the two-mode interaction induced by the EOM and the PDC. Thus, the optical loop and the PBS act in tandem to convert the two-mode EOM-PDC interaction between $\hat{a}_h$ and $\hat{a}_v$ into the interaction between cycling mode $\hat{a}_h$ and the emitted temporal modes $\{\hat{b}_j\}$. The state of emitted light is described in the next section.

## B. PROOF THAT SETUP GENERATES TN STATES

Our claim is that the light emitted from the optical loop is represented in the photon-number basis as a TN state of low bond dimension. To justify this claim, we consider the state of the vertically-polarized emitted light, while the horizontally polarized light continues to cycle in the optical loop. The intuition for the proof if that the entanglement between any two subsequent modes is meditated only by the cycling light, which places a limitation on entanglement between any two partitions of the emitted state. Figure 1 depicts this intuition and the structure of the generated TN states.

In the photon number basis, the transformations on the two polarization modes in the $i$-th cycle comprise a two-mode squeezing $U^{(i)}$ via PDC and an arbitrary linear transformation $V^{(i)}$ via EOM. The combined action of the PDC and the EOM on the two modes of light is represented by the operator

$$T^{(i)} \stackrel{\text{def}}{=} U^{(i)} V^{(i-1)} \equiv \sum_{k_i, \ell_i, \iota_i, j_i} T^{(i) j_i \iota_i}_{\ell_i k_i} \ket{\ell_i j_i} \bra{k_i \iota_i}, \quad (7)$$

acting on the cycling mode (with basis states indexed by $k_i$ or $\ell_i$) and the emission mode (indexed by $\iota_i$ and $j_i$), where we have defined $V^{(0)} = \mathbb{1}$. For reasonable pump powers, the relevant Hilbert space to be analyzed is restricted a small number $D-1$ of photons for a constant $D$. Thus, summation in Eq. (7) runs over $D$ values for each of the indices. Henceforth, we drop summation symbols over repeated indices.

During the first cycle, light in both the modes starts in the vacuum state $|00\rangle$. The two modes are entangled via the transformation to obtain the state

$$T^{(1)j_1\iota_1}_{\ell_1 k_1} |\ell_1 j_1\rangle \langle k_1 \iota_1 | 00 \rangle = T^{(1)j_1 0}_{\ell_1 0} |\ell_1 j_1\rangle, \quad (8)$$

where the two modes represent the horizontal cycling and the vertical emitted mode. Next, the light in the optical loop completes a cycle and interacts with the subsequent temporal mode at the EOM and at the PDC. Mathematically, an additional mode of light is appended to the system, and the light state is now represented by

$$T^{(1)j_1 0}_{\ell_1 0} |\ell_1 j_1\rangle \to T^{(1)j_1 0}_{\ell_1 0} |\ell_1 j_1 0\rangle \quad (9)$$

where the third mode represents the newly appended vertically polarized mode in vacuum state, and we have assumed that the cycling and emitted modes remain unchanged. After the interaction $T^{(2)}$ between the cycling light and the second emission mode, the state is

$$T^{(1)j_1 0}_{\ell_1 0} |\ell_1 j_1 0\rangle \xrightarrow{T^{(2)}} \left( T^{(2)j_2 \iota_2}_{\ell_2 k_2} |\ell_2 j_2 \rangle \langle k_2 \iota_2 | \right) T^{(1)j_1 0}_{\ell_1 0} |\ell_1 j_1 0\rangle$$
$$= T^{(2)j_2 0}_{\ell_2 \ell_1} T^{(1)j_1 0}_{\ell_1 0} |\ell_2 j_1 j_2\rangle, \quad (10)$$

where the first site is the cycling mode and the remaining sites refer to the emitted modes in the order of emission, a convention that we maintain in the remainder of this proof.

In general, after $m-1$ cycles, the state of light is

$$T^{(m-1)j_{m-1} 0}_{\ell_{m-1} \ell_{m-2}} \ldots T^{(2)j_2 0}_{\ell_2 \ell_1} T^{(1)j_1 0}_{\ell_1 0} |\ell_{m-1} j_1 j_2 \ldots j_{m-1}\rangle. \quad (11)$$

For simplicity, in the $m$-th cycle, we apply the swap operation on the cycling and the emitted light, which leads to the light in the cycling mode emitted from the setup and thus decouples the cycling mode from the emitted entangled state. Thus the state of the emitted $m$ modes is

$$T^{(m-1)j_{m-1} 0}_{\ell_{m-1} \ell_{m-2}} \ldots T^{(2)j_2 0}_{\ell_2 \ell_1} T^{(1)j_1 0}_{\ell_1 0} |0 j_1 j_2 \ldots j_{m-1} \ell_{m-1}\rangle, \quad (12)$$

where the cycling mode is in vacuum state and is decoupled from the emitted modes. Identically, the emitted state can be expressed as

$$T^{(m-1)j_{m-1} 0}_{j_m \ell_{m-2}} \ldots T^{(2)j_2 0}_{\ell_2 \ell_1} T^{(1)j_1 0}_{\ell_1 0} |j_1 j_2 \ldots j_{m-1} j_m\rangle. \quad (13)$$

The state (13) is the sum of basis states $|j_1 j_2 \ldots j_{m-1} j_m\rangle$ with coefficients in the form of a product of $D$-dimensional

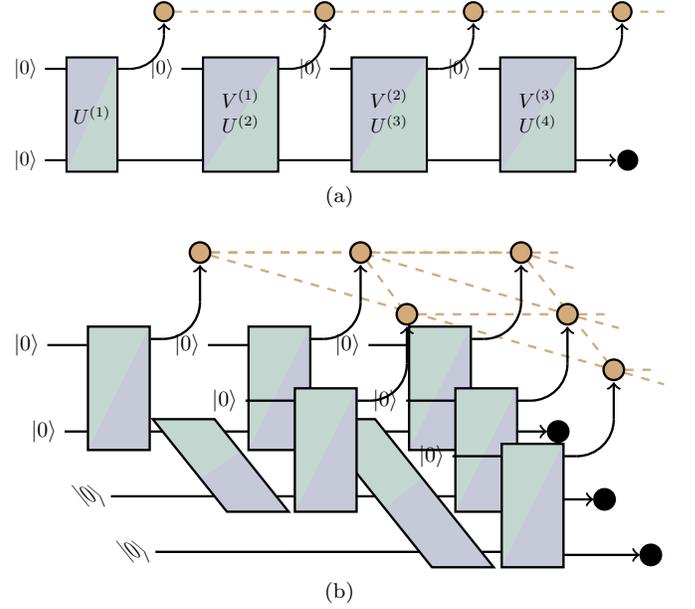

FIG. 1. **Quantum circuits to generate (a) one- and (b) 2D TN states**. The vertical blue-green rectangles represent EOM and PDC transformations between two polarization modes. The horizontal transformations represent the action of EOM, PDC and the fiber-loop on polarization and temporal modes (See text). The filled circles represent emitted modes while the filled black circled represent modes that continue to cycle in the loop. A dashed brown edge between two modes represents correlations introduced by the circuit between the modes. The overall structure of these modes and their connecting edges defines the structure of the TN. The labels in (b) are omitted for simplicity.

matrices $T^{(n)j_n 0}$; states in the form are matrix product states (MPS), or equivalently 1D TN state, of bond dimension $D$. This completes our proof. Furthermore, Eq. (13) also gives a general parameterization of the states that can be generated via our scheme.

### C. GENERATION OF $W$ AND GHZ STATES

While our TN state generation procedure can generate a broad class of TN states parameterized by the EOM parameters and the PDC interaction strength $\eta$ in each loop, we focus on two classes of entangled states for concreteness. Here we show how we can generate the $W$ and GHZ states using the scheme.

*Generation of W state* — Here we illustrate the generation of $W$ states, which admit an MPS representation with bond dimension $D = 2$ [6, 7]. We calculate in the single-rail basis and truncate our Hilbert space to no more than one photon in each mode, i.e., we neglect states with two or more photons in any of the modes. Equivalently, we ignore probabilities of order $|\eta|^4$ and higher powers, where $\eta$ is the PDC interaction parameter (5).

The generation of $W$ states can be accomplished by

setting $V^{(i)} = \mathbb{1}$ and $U^{(i)} = U$ over each cycle $i$ (i.e., by programming the EOM in the optical loop to leave the light unchanged, and by using the same pump pulse strength in each cycle) except the last cycle, in which the EOM effects a swap $V^{(m)} = \mathbb{S}$ and the pump is turned off $U^{(m+1)} = \mathbb{1}$. These parameters lead to an excitation $|1\rangle$ that is in a superposition over $m$ different sites.

For low pump strength, we can truncate the Taylor expansion of the unitary transformation (4) to low orders in $\eta$ as

$$\begin{aligned}
U &= \exp(\eta \hat{a}_h^\dagger \hat{a}_v^\dagger - \eta^* \hat{a}_h \hat{a}_v) \\
&\approx \mathbb{1} + (\eta \hat{a}_h^\dagger \hat{a}_v^\dagger - \eta^* \hat{a}_h \hat{a}_v) + \frac{1}{2}\Big(\eta^2 \hat{a}_h^{\dagger 2} \hat{a}_v^{\dagger 2} + \eta^{*2} \hat{a}_h^2 \hat{a}_v^2 \\
&\quad - |\eta|^2 (\hat{a}_h^\dagger \hat{a}_v^\dagger \hat{a}_h \hat{a}_v + \hat{a}_h \hat{a}_v \hat{a}_h^\dagger \hat{a}_v^\dagger)\Big) + \ldots \\
&= \mathbb{1} + (\eta \hat{a}_h^\dagger \hat{a}_v^\dagger - \eta^* \hat{a}_h \hat{a}_v) + \frac{1}{2}\Big(\eta^2 \hat{a}_h^{\dagger 2} \hat{a}_v^{\dagger 2} + \eta^{*2} \hat{a}_h^2 \hat{a}_v^2 \\
&\quad - |\eta|^2 (2\hat{a}_h^\dagger \hat{a}_v^\dagger \hat{a}_h \hat{a}_v + \hat{a}_h^\dagger \hat{a}_h + \hat{a}_v^\dagger \hat{a}_v + \mathbb{1})\Big) + \ldots,
\end{aligned} \quad (14)$$

which is correct up to the first two orders in $\eta$.

For calculations of the light state, we adopt the notation introduced in the main text where the first mode refers to the cycling horizontally polarized mode and the remaining modes are the emission modes arranged in increasing order of emission. For concreteness, we calculate the state of light at the end of four cycles. The starting state $|0000\ldots\rangle$ comprising vacuum in each of the modes is transformed by the PDC to

$$\begin{aligned}
|\psi_1\rangle &= U^{(1)} |0000\ldots\rangle \\
&= |0000\ldots\rangle + \eta |1100\ldots\rangle + \eta^2 |2200\ldots\rangle,
\end{aligned} \quad (15)$$

where the subscript in $|\psi_i\rangle$ denotes the number of completed cycles, and we have omitted the $(1-\eta^2)^{1/2}$ normalization factor for simplicity. Next, $U$ acts on the first and third modes while the remaining modes are left unchanged

$$\begin{aligned}
|\psi_2\rangle &= U^{(2)} |\psi_1\rangle \\
&= |0000\ldots\rangle + \eta |1010\ldots\rangle + \eta^2 |2020\ldots\rangle + \cdots \\
&\quad + \eta |1100\ldots\rangle + \sqrt{2}\eta^2 |2110\ldots\rangle + \cdots \\
&\quad + \eta^2 |2200\ldots\rangle + \cdots,
\end{aligned} \quad (16), (17)$$

where terms up to two orders in $\eta$ are considered. Then, $U$ acts on the first and fourth modes while the remaining modes are left unchanged

$$\begin{aligned}
|\psi_3\rangle &= U^{(3)} |\psi_2\rangle \\
&= |0000\ldots\rangle + \eta |1001\ldots\rangle + \eta^2 |2002\ldots\rangle + \cdots \\
&\quad + \eta |1010\ldots\rangle + \sqrt{2}\eta^2 |2011\ldots\rangle + \cdots \\
&\quad + \eta^2 |2020\ldots\rangle + \cdots \\
&\quad + \eta |1100\ldots\rangle + \sqrt{2}\eta^2 |2101\ldots\rangle \cdots \\
&\quad + \sqrt{2}\eta^2 |2110\ldots\rangle + \cdots \\
&\quad + \eta^2 |2200\ldots\rangle + \cdots.
\end{aligned} \quad (18), (19)$$

The terms of each order in $\eta$ can be grouped to obtain

$$\begin{aligned}
|\psi_3\rangle &= |00\ldots\rangle + \eta |1\rangle \otimes (|001\ldots\rangle + |010\ldots\rangle + |100\ldots\rangle) \\
&\quad + \eta^2 |2\rangle \otimes (|200\ldots\rangle + |020\ldots\rangle + |002\ldots\rangle) + \\
&\quad + \sqrt{2}\eta^2 |2\rangle \otimes (|011\ldots\rangle + |101\ldots\rangle + |110\ldots\rangle) \\
&\quad + O(\eta^3)
\end{aligned} \quad (20)$$

In the last cycle, the EOM performs the swap operation and the pump is turned off, thereby allowing the light in the cycling mode to exit the loop. Thus, the final state is given by

$$|\psi_4\rangle = |0\rangle \otimes |000\rangle + \eta |1\rangle \otimes \big(|001\rangle + |010\rangle + |100\rangle\big), \quad (21)$$

which has been truncated to first order in $\eta \ll 1$. Writing the output state in the order of the departure of the photons, we have the unnormalized state

$$|\tilde{\psi}_4\rangle = |000\rangle \otimes |0\rangle + \eta \big(|001\rangle + |010\rangle + |100\rangle\big) \otimes |1\rangle, \quad (22)$$

which is interpreted a heralded $W$ state as follows. If no photon is emitted out of the loop in the last cycle, (i.e., the last departure time mode), then only vacuum was be emitted in the preceding three modes (first three cycle). On the other hand, with probability $\eta^2$, a single photon will be emitted in the last cycle, and this emission 'heralds' (or announces belatedly) the emission of a $W$ state

$$|\psi\rangle_W = |001\rangle + |010\rangle + |100\rangle \quad (23)$$

in the preceding three modes. Fig. 2a depicts the quantum circuit representing this scheme.

In general, running the experiment for $m+1$ cycles will produce a heralded $m$-mode $W$ state

$$|\psi\rangle_W = |00\ldots 01\rangle + |00\ldots 10\rangle + \cdots + |10\ldots 00\rangle. \quad (24)$$

Note the heralding photon is emitted after each of the other modes, so experiments that require the $W$ state can employ post-selection to choose only that subset of data in which a photon is emitted in the last cycle. The relative weights of the different components in the state of Eq. (1) can be adjusted by varying the strength of the pumping and thereby the interaction strength $\eta_i$ in each cycle $i$.

*Generation of four-mode GHZ state* — Another important TN state is the GHZ state. We show here that the setup can be used to generate a four-mode GHZ state. The four-mode GHZ state is given by $|0000\rangle + |1111\rangle$ or equivalently by $|0011\rangle + |1100\rangle$. To obtain this state, the pump and the EOM are turned on in the first and third cycle and turned off in the second and fourth cycle, i.e., $U^{(i)} = V^{(i)} = \mathbb{1}$ for even $i$ and $U^{(i)} = U, V^{(i)} = \$$ for odd $n$. This creates two excitations in a superposition over the first two and and the last two temporal modes, which is the required diluted GHZ state.

Similar to the $W$ state analysis, we calculate in the ordering of arrival time and ultimately switch to departure-time ordering for analyzing the output. The action of the first unitary on the vacuum state gives the unnormalized state

$$|\psi_1\rangle = U^{(1)}|0000\rangle$$
$$= |0000\rangle + \eta|1100\rangle + \eta^2|2200\rangle + \ldots. \quad (25)$$

The first EOM operation is set such that it swaps the state of light in its two modes

$$V^{(1)} = \$ \stackrel{\text{def}}{=} \begin{pmatrix} 0 & 1 \\ 1 & 0 \end{pmatrix}, \quad (26)$$

which gives

$$|\psi_{2'}\rangle = V^{(1)}|\psi_1\rangle \quad (27)$$
$$= |0000\rangle + \eta|0110\rangle + \eta^2|0220\rangle + \ldots. \quad (28)$$

Next, the light enters the PDC whose pump power is set to zero ($U^{(2)} = \mathbb{1}$) and the EOM action is set as identity $V^{(2)} = \mathbb{1}$. Hence, the state is unchanged $|\psi_{3'}\rangle = |\psi_{2'}\rangle$ on going through the PDC and the EOM. The vertical polarization mode is coupled out of the optical loop via the PBS; the loop elements will subsequently act on the next temporal mode. Upon action $U^{(3)} = U$ of the PDC in the next cycle, the state changes to

$$|\psi_3\rangle = U^{(3)}|\psi_{3'}\rangle = U^{(3)}|\psi_{2'}\rangle$$
$$= |0000\rangle + \eta|1001\rangle + \eta^2|2002\rangle$$
$$+ \eta|0110\rangle + \eta^2|1111\rangle + \eta^2|0220\rangle + \ldots. \quad (29)$$

Ignoring terms containing amplitude $\eta^2$ and higher powers of $\eta$, we obtain

$$|\psi_3\rangle = |0000\rangle + \eta\left(|1001\rangle + |0110\rangle\right). \quad (30)$$

As usual, a swap operation at the end of the process leads to the emission of the cycling mode. Arranging the modes in order of departure, the output light is represented by the state

$$|\psi_3\rangle = |0000\rangle + \eta\left(|0011\rangle + |1100\rangle\right), \quad (31)$$

which is a superposition of the vacuum state with the GHZ state. The relative weights of the two terms can be adjusted by varying the pumping strength. This completes our construction of GHZ-type states. We depict the circuit corresponding to this construction in Fig. 2b.

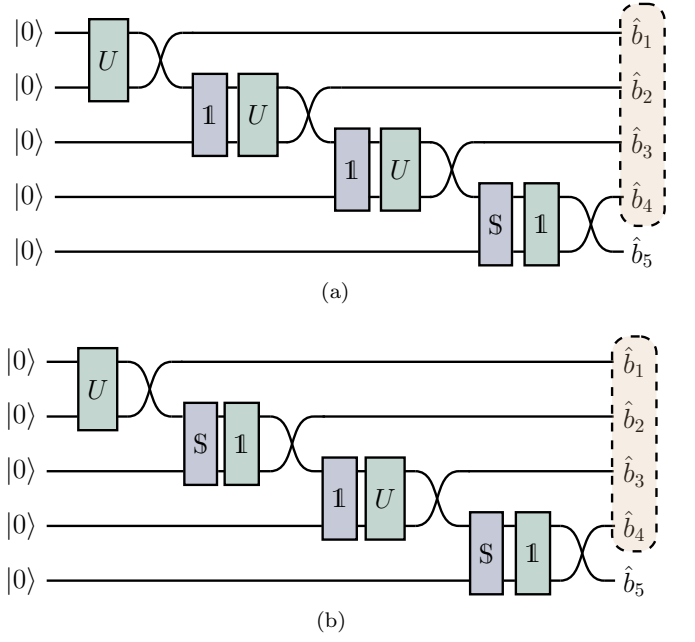

FIG. 2. **(a) The circuit for generating the heralded three-mode $W$ state:** The EOM only effects identity transformations for the first three cycles, and performs the swap operation (labelled $\$$) in the fourth cycle. The PDC is pumped with identical power for the first three cycle and in the last cycle, the pump is turned off. The resultant state in the modes corresponding to $\hat{b}_1, \ldots, \hat{b}_4$ is the heralded three-mode $W$ state. **(b) The circuit for the generation of the GHZ state.** The state of light in modes $\hat{b}_1, \ldots, \hat{b}_4$ is a four-mode GHZ states.

### D. SIMULATIONS OF EXPERIMENT

Here we present relevant details of the simulations for the results presented in Figs. 4 and 5. The state of light in the setup was represented in the photon number basis using a 1D TN, i.e., a MPS representation because a full density matrix representation of the generated output was not feasible. We used the library mpnum [8] to simulate 1D TNs and basic operations thereon. The Fock space was truncated to no more than three photons in each mode. We constructed the PDC and EOM operators in the photon-number basis representation via python package qutip [9] and translated these into a matrix product operator (MPO) form. Similarly, measurements were represented as projection MPOs.

We include the experimental imperfections of loss, mode-mismatch and a finite number of measurements. Loss and mode-mismatch are modeled by assuming an additional beamsplitter in the loop. This beamsplitter is assumed to receive vacuum in one input port and the cycling mode in the other, and one of its output ports is grounded while the other continues as the new lossy cycling mode. Experimental noise is simulated by adding the expectation value calculated from the simulations with

Gaussian random noise, i.e., Gaussian random numbers that have mean zero and variance equal to the operator variance divided by the inverse of the number of measurements.

The simulations for the quantum variational algorithm (Fig. 5 of main text) are performed as follows. A random set of circuit parameters is chosen as the initial setting of the circuit. Next, measurements are simulated on the state of the system based on the given Hamiltonian. Variational minimization is performed over the circuit parameters to minimize the energy with respect to the given Hamiltonian. The fidelity of the state is computed against the exact target state, which is also represented as an MPS.

Simulations for feedback control (Fig. 4 of main text) start with the ideal circuit parameters, i.e., parameters that lead to a correct state generation without loss and other imperfections. Imperfect state generation is simulated as described above. Next, circuit parameters that maximize the fidelity with respect to a given target state are found by variational maximization. The fidelity of the state thus generated is higher by one to two orders of magnitude higher than those obtained without variational optimization. In both these cases, we use the constrained optimization by linear approximation (COBYLA) algorithm to perform constrained numerical optimization [10, 11].

### E. EXPERIMENTAL CONSIDERATIONS

Here we discuss the effect of experimental imperfections in the implementation of the all-optical generation of TN states using the setup depicted in Figs. 1 and 3 of the main text. We analyze the effect of losses, higher-order photon emissions and dark counts on the fidelity of the generated states and we estimate the rates at which 1D and 2D TN states can be generated using the setup.

First, we consider the effect of losses. We distinguish between two kinds of losses in the setup. The first kind of loss that acts on the cycling mode, i.e., only degrades the quality of the cycling horizontally polarized light. This kind of loss can arise from three sources: (i) losses in the optical loop, which are small for free-space loops, but can be considerable for fiber-loops because of the requirements of in- and out-coupling and of polarization compensation, (ii) imperfect transmission at the nonlinearity, and (iii) mode-mismatch at the nonlinearity, which could arise due to a difference between input and output spatial modes of a PDC waveguide if a waveguide-based nonlinearity is used. This mode-mismatch can be modeled as a loss term if one of the two incoming polarization modes is in the vacuum state. The second kind of losses are incurred at the out-coupling of the light from the loop. This might result from imperfect coupling from free-space to fiber if a free-space loop is used, and also from imperfect detection efficiencies if detection is included in the setup.

The losses degrade the quality of the generated states. We begin by considering the losses incurred by the state at the out-coupling from the optical loop. These losses act identically on each of the $m$ emitted modes. Furthermore, we define $\epsilon_1$ as the maximum number of excitations ('|1⟩'s) in any of the terms in the superposition of the generated states. For instance, for the heralded W- and diluted GHZ-states, $\epsilon_1 = 2$. As we are working in the photon-number basis, $\epsilon_1$ is an upper bound on the number of photons in any term of the superposition. The fidelity of the generated state with respect to the target state depends only on the total number of photons in the modes and not on the number of modes itself. Explicitly, the fidelity scales no worse than $(\mathcal{T}_{\text{out}})^{\epsilon_1}$ for out-coupling transmission probability $\mathcal{T}_{\text{out}} \leq 1$ and does not depend on the number $m$ of emitted modes. This also means that the out-coupling losses are independent of the number of loops in the setup, and thus independent of the dimension of the generated TN state.

Next, we consider the effect of the cycling losses, i.e., the losses incurred by the light that is cycling in the loop. The effect of these losses becomes more detrimental with increasing size of the generated state since the cycling mode incurs these losses repeatedly in each cycle. Although the exact scaling of the state fidelity will depend on the interaction between the cycling and the emitted mode, we provide heuristic estimates. We define $\epsilon_2$ as the maximum number of photons in the cycling mode during the state generation and $\mathcal{T}_{\text{cyc}}$ as the transmission probability for the cycling mode. For a generated state over $m$ modes, the fidelity is expected to scale no worse than $(\mathcal{T}_{\text{cyc}})^{m\epsilon_2}$. Thus, the effect of the cycling losses increases quickly with the number of modes of light emitted. Another possible imperfection arises from the transmission probability $\mathcal{T}_{\text{cyc}}$, which might decrease with additional loops added to the setup but this imperfection can be made very small; for instance transmission probabilities up to 98% can be reached using a free-space loop with wavelength-optimized coatings of all components [12]. Thus, other than the minor reduction in transmission probability, the scaling of generated state fidelity is expected to be independent of the dimension of the generated TN state. This situation is this more favorable that the case of classical computation of TN states, wherein the computational hardness depends strongly on the dimensionality of the considered state.

Finally, we note that the above analysis of fidelity only provides pessimistic estimates since states with the incorrect number of photons can often be discarded via post-selection. In other words, states that have incurred photon loss can be eliminated from the analysis, thereby increasing the effective fidelity of the experiment.

Another important experimental consideration is source efficiency, which depends on its brightness and the coupling efficiency of the generated modes to the loop. With the current technology, PDC source brightness levels of

up of $3.8 \cdot 10^{11}$ photon pairs/(W·s·m) have been demonstrated [13]. Current laser technology allows for performing MHz-rate experimental runs, where each experiment can involve up to 80 coherent laser pulses. Taking into account a free space loop transmission of 98%, a mode overlap of 90% and a coupling efficiency (between loop and nonlinearity) of 60% per cycle, a detection efficiency of 85% and a sample length of 8mm, the generation rate of a four-mode $W$- or GHZ- state is $\approx 2 \cdot 10^5$ states per second and per mW of pump power. For reasonable pump powers of $10\mu$W, $\approx 2000$ states per second are generated. Similarly, generation of five-mode 1D TN states could be accomplished at a rate of $2000 \cdot 60\% \approx 1200$ Hz. We can estimate the generation rates for 2D TN states by following a similar logic, where we consider imperfect coupling (between two nonlinearities) of 60% twice in the loop and an additional 98% transmission arising from the additional free-space loop. With these experimental parameters, we can obtain a fifteen-mode (five cycles) 2D TN state with a rate of $\approx$ 84 Hz.

Now we consider the effect of the higher-order photon emission terms on the generates states. The experimental parameters described above correspond to the PDC parameter value of up to $|\eta| \approx 0.1$ or equivalently 0.01 probability of the emission of one photon pair. Thus, the higher-order emission probability in each run of the experiment is around $|\eta|^4 \approx 10^{-4}$ for the emission of two pairs of photons. This emission rate will lead to $\approx 20$ events per second in which an incorrect state is generated out of the $\approx 2000$ events in which a state, correct or incorrect, is heralded. These false events can be identified and discarded by performing moderate number-resolved detection, which is possible either via time-demultiplexing or via superconducting transition-edge sensors [14–17]. Thus, higher-order emissions from the PDC source do not have an appreciable detrimental effect on the experimental rates or outcomes.

Another possible source of imperfections is dark counts at the detectors, i.e., events in which the detector clicks without the concomitant incidence of a photon. Dark counts rates are very small ($\approx$100 Hz) for modern nanowire detectors [18]. Moreover, precise incidence times of the pumped photons can be measured and detector dark counts occur randomly with uniform probability over time. This allows discarding all the dark counts that arise at times other those time at which the photon pulses are expected. Therefore the effective dark count rates come down to 1Hz, which is negligible for the current proposal.

In summary, the experimental generation of the TN states is possible with reasonable state-generation rates under current experimental losses. Other sources of imperfection, namely higher-order emissions and dark counts do not have any appreciable impact on the quality or rates of state generation.



\end{document}